# Normalization and correction factors for magnetic tunnel junction sensor performances comparison

E. Monteblanco, A. Solignac, C. Chopin, J. Moulin, P. Belliot, N. Belin, P. Campiglio, C. Fermon, M. Pannetier-Lecoeur

*Abstract*— **In this manuscript we propose a calculation method where the magneto-resistive elements are modelled as fluctuating resistances to correct the output voltage noise of magnetic tunnel junction (MTJ) from standard electronic circuits. This method is validated on single elements, partial and full Wheatstone bridge circuits, giving rise to a correction factor affecting the output voltage noise as well as sensitivity values. Combining the correction factor and a normalization by the number of MTJs pillars and the pillar surface, we show that the performances extracted by this method allow universal comparison between any results from literature.**

## I. Introduction

Magnetic tunnel junctions (MTJ) play an important role on current technological devices, such as non-volatile memories (MRAM) [1], spintronic nanosillators (STO) [2], neuromorphic computing [3], magnetic sensing for industrial applications like automotive or mobile phones [4]. Even more, it is a promising candidate for the development of ultrasensitive magnetic field sensors capable to detect extremely weak magnetic fields, such as biosensors or for biomedical applications [5,6], magnetic resonance imaging (MRI) or magnetoencephalography (MEG) [7].

MTJ are based on spin electronics and are composed schematically of two magnetic layers separated by a tunnel barrier through which spin polarized electrons are transported. The tunnel magnetoresistance (TMR) ratio is in the order of 200% at room temperature, leading to high output voltages. However, the overall performances for sensing purpose include the evaluation of the noise of the MTJ elements. The noise provided by TMR sensors comes from the contributions of several sources such as the thermal noise, shot noise, low-frequency noise (1/f) and random telegraphic noise (RTN) [8].

The performance of these type of sensors based on the TMR effect is mainly limited by the high 1/f noise level presents in the low frequency range. It might be induced by the presence of magnetic domain fluctuations on magnetic electrodes (free layer domains) and also by defects inside the insulator barrier (traps, oxygen vacancies, etc.) [8-11]. A lot of noise and performances studies on TMR sensors have been reported [12] but are somehow difficult to compare due to different electronic circuits and normalization. In this paper we first introduce a calculation method for standard electronic circuit correction (single element, ¼ and full Wheatstone bridge) including various normalization factors (bias voltage, size, number of pillars). Then the method is validated on several devices with different sizes and number of MTJ pillars.

## II. Calculation framework

If a TMR sensor with a resistance $R$ and a transversal area $A$ is biased by a bias voltage $V_{TMR}$, its noise spectral density $S_V[V^2/Hz]$ is composed of two contributions: the white noise (thermal noise $S_{V,th}$ in the condition $qV \ll k_B T$, where $q$ is the electron charge, $k_B$ the Boltzmann constant and $T$ the temperature) and the 1/f low frequency noise $S_{V,1/f}$ [8]. The total noise spectral density is expressed as $S_V = S_{V,th} + S_{V,1/f}$. When the bias voltage increases, the white noise includes the shot noise term $S_{V,shot} = 2qI$ induced by the individual electrons of the current I passing through the tunnel barrier. Random telegraphic noise (RTN) could also be present. The equation of each contribution is given in table 1

Noise measurements can be performed using different types of electronic circuitry [8] such as a single element [13], a ¼

This work was supported in part by the French ANR under Projects AdvTMR (ANR-18-ASTR-0023), NeuroTMR (ANR-17-CE19-0021) and CARAMEL (ANR-18-CE42-0001).

E. Monteblanco is working at SPEC (CEA Saclay, CNRS, Université Paris-Saclay) France. Email: noel.montblanc@cea.fr

A. Solignac is working at SPEC (CEA Saclay, CNRS, Université Paris-Saclay) France. Email: aurelie.solignac@cea.fr

C. Chopin is working at SPEC (CEA Saclay, CNRS, Université Paris-Saclay) France. Email: chloe.chopin@cea.fr

J. Moulin was working at SPEC (CEA Saclay, CNRS, Université Paris-Saclay) France. He is now working at INSTN (CEA Cadarache) Email: julien.moulin@cea.fr

P. Belliot is working at (CrivaSense Technologies SAS, 91190 Saint Aubin) France. Email: pierre.belliot@crivasense.com

N. Belin is working at (CrivaSense Technologies SAS, 91190 Saint Aubin) France. Email: noemie.belin@crivasense.com

P. Campiglio is working at (CrivaSense Technologies SAS, 91190 Saint Aubin) France. Email: paolo.campiglio@crivasense.com

C. Fermon is working at SPEC (CEA Saclay, CNRS, Université Paris-Saclay) France. Email: claude.fermon@cea.fr

M. Pannetier-Lecoeur is working at SPEC (CEA Saclay, CNRS, Université Paris-Saclay) France. Email: myriam.pannetier-lecoeur@cea.fr

TABLE 1
Electronic circuits modelled, noise equations and correction factors

| | (a) | (b) | (c) |
|---|---|---|---|
| Electronic circuit | 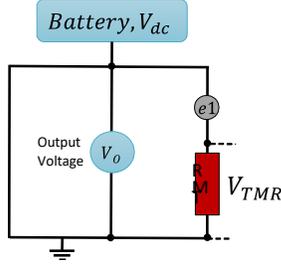 | 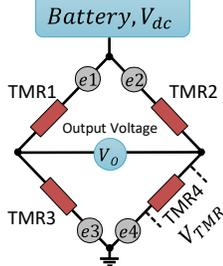 | 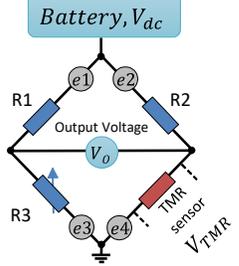 |
| Equations | | Correction factor (CF) | |
| Thermal noise | $S_{V,th} = 4k_B T R$ | 1 | 1 | $2\left(\dfrac{M-1}{M}\right)$ |
| Shot noise | $S_{V,shot} = 2qI$ | 1 | 1 | $\left(\dfrac{M-1}{M}\right)^2$ |
| 1/f noise | $S_{V,1/f} = \dfrac{\alpha V_{TMR}^2}{NAf}$ | 1 | 1 | $\left(\dfrac{M-1}{M}\right)^2$ |
| Hooge parameter | $\alpha = \dfrac{S_{V,1/f}\cdot NA\cdot f}{V_{TMR}^2}$ | 1 | 1 | $\left(\dfrac{M-1}{M}\right)^2$ |

T corresponds to the temperature, f is the frequency, $k_B$ to the Boltzmann constant, R is TMR sensor resistance, I the applied current, q the electron charge, N is the number of pillars, A is the pillar surface, $V_{TMR}$ corresponds to the TMR sensor applied bias voltage and $M = \dfrac{R1+R3}{R3} = \dfrac{R2+R4}{R4}$. The presence of RTN would lead to an additional term in the total noise, on which the CF should be applied.

Wheatstone bridge [14] or a complete Wheatstone bridge [4], see figures (a), (b) and (c) on table 1; the latter is often used in industrial sensor applications for eliminating drifts such as temperature drift. The advantage of the Wheatstone bridge is that it removes the DC component and shifts the signal of interest around zero volt. The amplification of the field-dependent signal is then easier and can be more important, and the magnetic field sensor power supply can also be larger without saturating the amplifiers. However two elements over the four should be exposed to the field to detect or should have an opposite sensitivity direction [4,8]. The ¼ bridge consists of a TMR sensor, a variable resistor to balance the sensor resistance and thus the bridge and two higher value resistors that stabilize the current in the bridge. The ¼ bridge is an easy circuit to work on in a laboratory environment but it is less compatible with multi TMR sensors measurements and most industrial applications where balancing the variable resistor at each TMR sensor is not possible. A full bridge with four identical resistors solves the problem while maintaining the interest of the bridge and eliminating drifts such as temperature drifts. Since the design of the electronic circuit is affecting the extracted noise output values, a correction factor should be taken into account. In the following, we present a simple method to get the correction factor to compute independently from the measurement configuration the Hooge-like parameter of the MTJ, sensitivity and the output noise.

A TMR sensor can be represented by an equivalent electronic circuit composed by one resistance R and a small fluctuation responsible of noise. The noise can have different origin: resistance fluctuations, current fluctuations or voltage fluctuations. Regardless of its origin, it could be summarized as a resistance $R$ with a fluctuator $e = \sqrt{S_V}$, as depicted in figures on table 1. For this method, the voltage source is supposed to be stable and thus its noise is neglected. By considering the general case of a Wheatstone bridge composed by four elements of resistance $R_i$ in series with small fluctuators $e_i$ and from small signal analysis, non-null resistances and the Millman theorem lead to the following equation for the output voltage of the bridge:

$$Output_{voltage} = \frac{e1\times R3}{R1+R3} + \frac{e3\times R1}{R1+R3} - \frac{e2\times R4}{R2+R4} - \frac{e4\times R2}{R2+R4} \quad (1)$$

This relationship impacts directly to the output noise spectral density of the bridge which is the square of the output voltage. As the four sources of noise $e_i$ are considered not correlated, the diagonal terms of the equation are suppressed and the output noise can be expressed as:

$$S_v\left[\frac{V^2}{Hz}\right] = output_{voltage}^2 = e1^2 \times \left(\frac{R3}{R1+R3}\right)^2 + e3^2 \times \left(\frac{R1}{R1+R3}\right)^2 + e2^2 \times \left(\frac{R4}{R2+R4}\right)^2 + e4^2 \times \left(\frac{R2}{R2+R4}\right)^2 \quad (2)$$

Assuming the TMR sensor as the only source of shot noise and main source of 1/f noise, then the output noise could be expressed by the equations given in table 1 for the various electronic circuits. While the noise is identical in a single TMR sensor circuit and in a full bridge by considering the voltage $V_{TMR}$ as the bias voltage across the TMR sensor, for the ¼ bridge, the correction factor (CF), appears with $M = \frac{R1+R3}{R3} = \frac{R2+R4}{R4}$. It should be removed (divided) from each noise contribution (see equations in table 1) in order to obtain the real output noise from the TMR sensor. A consequence of this correction factor is the need to consider the variation of the TMR sensor resistance versus field and versus voltage, as it will be seen in the experimental section of the paper.

TABLE 2
SUMMARY OF DEVICES PARAMETERS AND MEASUREMENT CONFIGURATIONS

| Device | #pillars (N) | Diameter (μm) | $R_0$ (kΩ) | $TMR_{0,1mA}$ (%) | $R_0 \cdot A/N$ (kΩμm²) | Configuration |
|---|---|---|---|---|---|---|
| 1 | 4 | 5 | 2.75 | 167 | 13.5 | ¼ Wheatstone bridge |
| 3 | 10 | 5 | 6.75 | 138 | 13.2 | ¼ Wheatstone bridge |
| 4 | 2 | 3 | 3.50 | 150 | 12.4 | ¼ Wheatstone bridge |
| 5 | 30 | 5 | 18 | 150 | 11.8 | ¼ Wheatstone bridge |
| 7 | 2 | 15 | 0.16 | 172 | 14.3 | ¼ Wheatstone bridge |
| 8 | 2 | 10 | 0.47 | 175 | 18.6 | ¼ Wheatstone bridge/Single element |
| 9 | 30 | 3 | 45 | 210 | 10.6 | Full Wheatstone bridge |
| 10 | 30 | 15 | 1,9 | 200 | 11.5 | Full Wheatstone bridge |

## III. EXPERIMENTAL RESULTS

The multilayer stack of the TMR sensor studied in this manuscript has the following structure (see Figure 1 (a)) to linearize the sensor response [15,16]: SiO$_2$ / Ta(5) / SyF1 / MgO / SyF2 / Ta(10) (in brackets the thickness are in nanometers). The synthetic ferrimagnet SyF1 is composed by a multilayer PtMn(25) / CoFe(2.3) / Ru(0.83) / CoFe(1) / Ta(0.1) / CoFeB(1.5) with a strong RKKY coupling and corresponds to the reference layer which has a fixed magnetization under the field to detect. The SyF2 is composed by CoFeB(1.5) / Ta(0.1) / CoFe(1) / Ru(2.6) / CoFe(2) / PtMn(16) with a softer RKKY coupling and corresponds to the "free" layer which magnetization rotates with the field to detect. The samples were first post annealed at 300 °C during 30 min with a magnetic field of 1 T to pin the hard SyF1 magnetization into the sensitivity axis direction (0°) and to obtain a good crystallization of the CoFeB barrier interfaces. A second annealing was done at 300 °C during 30 min with a magnetic field of 80 mT to fix the SyF2 layer magnetization at 90 ° from the reference sensitivity axis. Samples were processes after the annealing using standard optical UV lithography and by ion milling. The diameter of the MTJs pillars was between 3-15 μm with a RA product at 0 mT typically around 12-20 kΩ.μm² (see table 2). Samples were wire bonded to a chip for the transport and noise measurements at room temperature. The TMR for all devices yields 200% at room temperature, measured at 100 mV, an evidence of the good quality of the MgO barrier and the correct filtering of electrons [17]. The orthogonal magnetization positions of both SyFs allows to linearize the magnetic response of the TMR when the magnetic field is applied on the sensitivity direction as can been seen in figure 2(a) around zero field. Parallel (P) and antiparallel (AP) magnetic configuration [14] is translated as lower and higher resistance in this curve and the SyF1 spin-flop regime appears overcoming positive 50 mT. Geometrical and structural parameters have been optimized to reduce the hysteresis of the curve, remaining linear around zero field, as it is expected by sensors applications.

TMR sensors studied in this paper exhibit a variable number of pillars of various sizes, as depicted in table 2. In order to remain below the breakdown voltage or to avoid a possible degradation of the MgO barrier, the maximum $V_{pillar}$ applied to the TMR sensors was around 1 V. In order to characterize correctly the 1/f noise from the TMR sensor, the setup is installed in a mu-metal magnetically shielded room. A battery is used to provide a stable bias voltage to the devices through a balanced Wheatstone bridge (1/4 or complete bridge) or through the single element (in series with a high resistance to avoid shortcuts). The output signal of the bridge is amplified using an INA103 low-pass amplifier (gain=489) and then amplified again (gain=10) and bandpass filtered (0.1-3 kHz). For single element, the DC voltage is AC coupled using a band-pass filter (SR360) and no DC amplification could be used, because of the high DC component. Temporal signals are acquired by an acquisition card. A fast Fourier transform (FFT) is used to measure the noise spectral density (average of 30 times). Since the sensitivity is defined as the output voltage per unit field divided by the bias voltage, we use a complementary technique to measure the sensitivity of our devices. An alternative external calibrated magnetic field ($H_{850nT}^{rms}$)) of 850 nTrms at 30 Hz was applied along the sensitivity axis, so using the output signal at 30 Hz ($V_{30Hz}^{rms}$), we can calibrate the sensor. From the output noise voltage, we define the sensitivity and the detection limit as follows:

$$Sensitivity\ \left[\frac{\%}{T}\right] = \frac{100.V_{30Hz}^{rms}}{H_{850nT}^{rms}V_{TMR}} \quad (4)$$

$$Detection\ limit\ \left[\frac{T}{\sqrt{Hz}}\right] = \frac{\sqrt{S_V}}{Sensitivity.V_{TMR}} = \sqrt{\frac{\alpha}{NAf}}\frac{H_{850nT}^{rms}.V_{TMR}}{V_{30Hz}^{rms}} \quad (5)$$

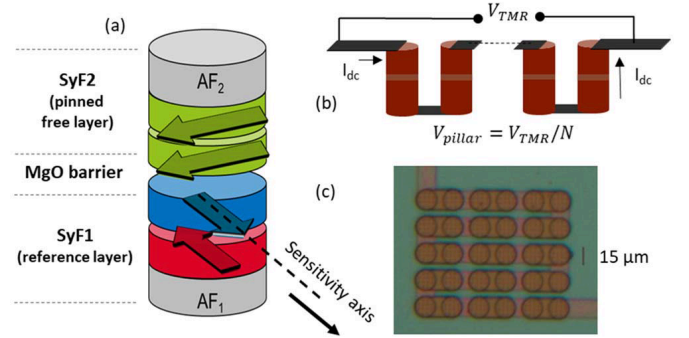

Fig. 1. (a) Principle schematic of the TMR stack. (b) TMR sensor element composed by $N$ pillars connected in series. (c) Optical image of the TMR sensor element (30 pillars of diameter 15μm) after process.

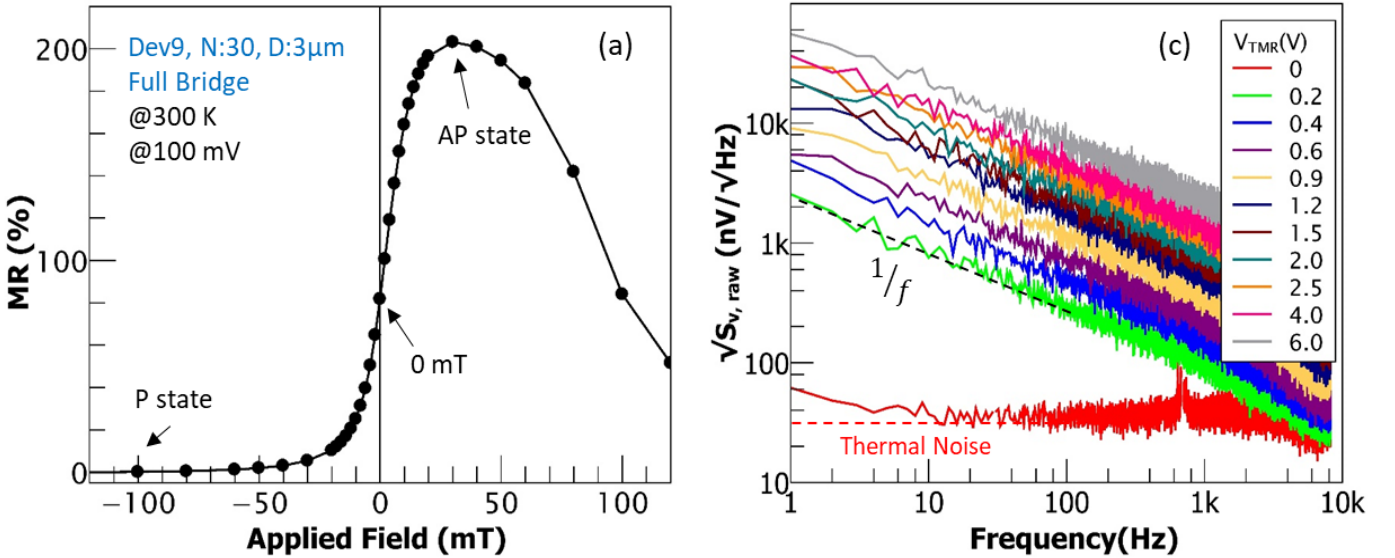

Fig. 2. (a) Magnetoresistance curve as a function of the applied field and (b) Output voltage noise as a function of the frequency for device 9 (full bridge configuration) and as function of the voltage bias.

In this work, the voltage noise was measured as a function of the frequency at 0 mT for different $V_{TMR}$ values, using the single element, ¼ and complete Wheatstone bridge configuration and for the different sensors listed in table 2 in order to validate the method developed in part II. For instance, the output voltage $\sqrt{S_V}$ on figure 2(b) corresponds to the complete Wheatstone bridge configuration of device 9. The red dashed line corresponds to the thermal noise $\sqrt{S_{V,th}}$ which varies from 3 to 30 nV/√Hz at room temperature depending to the TMR sensor resistances. A dashed black line is introduced to indicate the 1/f noise contribution. There is no signature of random telegraph noise (RTN) in our devices [16] so, the 1/f low-frequency noise will be dominant for TMR sensors.

As described in table 2 and figure 1(b), arrays of $N$ MTJs pillars in series with different diameters A are fabricated and measured. Arrays of TMR in series are generally used to reduce the applied voltage to each pillar ($V_{pillar} = V_{TMR}/N$), to keep a high TMR ratio and to reduce the 1/f noise. The surface of the pillars can help to tune the resistance obtained as well as the footprint. Clearly, TMR sensor performances are also affected by these geometrical parameters and a normalization factor is applied to obtain performances for a single pillar of 1μm² area and it will be detail in the following.

Figure 3(a) shows the output voltage noise $\sqrt{S_v}$ at two frequencies (10 Hz and 1 kHz) as a function of the bias voltage for devices 1 to 9 measured at 0 mT in various configurations, single element, ¼ and full Wheatstone bridge. We observe how the noise increases linearly with the voltage independently of the frequency until it reaches a saturation regime at high voltages [18]. From the 1/f equation on table 1, we can obtain a relation to compare the output noise from different devices $\sqrt{S_{v,N_i}} = \frac{\sqrt{N_iA_j}}{\sqrt{N_jA_i}}\sqrt{S_{v,N_j}}$ taking into account the same $V_{pillar}$. The $\sqrt{S_v}$ normalized (divided by the factor $\sqrt{N/A}$) and corrected by one pillar are introduced on figure 3(a). It shows the same voltage values and behavior, evidence of the good performance of the device as well as of the method for corrections and normalization.

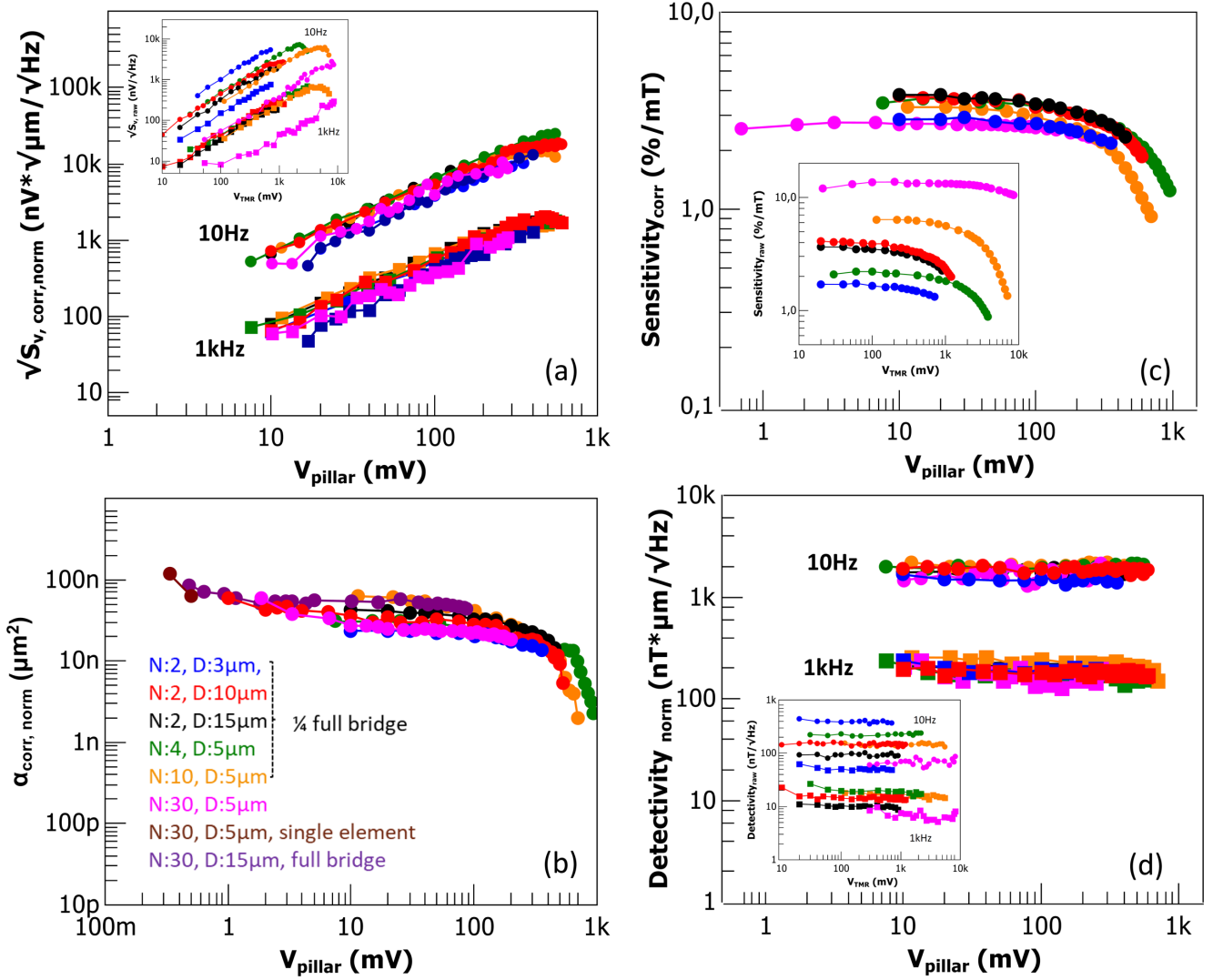

Fig.3. Noise, sensitivity and detectivity sensors performance for all devices listed in table 2. (a) Output voltage noise corrected and normalized per pillar as a function of the applied voltage for 0 mT at 10 Hz and 1 kHz. Inset: raw data. (b) Corrected and normalized Hooge-like parameter as a function of the applied voltage normalized per pillar at 0 mT. (c) Sensitivity corrected as a function of the applied voltage on a pillar at 0 mT. Inset: raw data. (d) Detection limit normalized as a function of the applied voltage on a pillar (at 10 Hz and 1 kHz) at 0 mT. Inset: raw data.

As a difference with other studies [19, 20], the correction factor (table 1) is included to obtain the real TMR sensor performances, especially through the factor $M$ which depends on the TMR sensor resistance in the state of measurement. For this purpose, the variation of the TMR sensor resistance versus field and voltage is measured. Also, $\alpha$ parameters are normalized taking into account the number of pillars and effective area, for various devices.

TABLE 3
SUMMARY OF CORRECTION AND NORMALIZATION FACTORS (INTO A SINGLE PILLAR OF AREA 1 μm²)

|  |  | Equations | Correction | Normalization into a single pillar of area 1 μm² |
|---|---|---|---|---|
| Noise | $V/\mu m/\sqrt{Hz}$ | $\sqrt{S_v}$ | $\sqrt{CF}$ | $\sqrt{N/A}$ |
| Sensitivity | $\%/mT$ | $\dfrac{100 \cdot V_{30Hz}^{rms}}{H_{850nT}^{rms} V_{TMR}}$ (4) | $\sqrt{CF}$ | 1 |
| Detectivity | $T/\mu m/\sqrt{Hz}$ | $\sqrt{\dfrac{\alpha}{NAf}} \dfrac{H_{850nT}^{rms} \cdot V_{TMR}}{V_{30Hz}^{rms}}$ (5) | 1 | $1/\sqrt{A \cdot N}$ |

The method provides the correct quantification of the 1/f noise as a good agreement of α parameter is observed for all the electronic configurations as well as for all the different TMR sensor parameters (size, number of pillars). The non-constant alpha parameter versus bias voltage is unexpected but it will not be discussed in this paper [18].

Figure 3(c) shows the sensitivities of the TMR sensors with various sizes and number of pillars, as a function of the applied voltage at 0 mT and measured with different electronic circuit. In all cases sensitivity decreases, following the same trend as the TMR [18]. At 0 mT and low bias voltage sensitivities reach values between 3-4 %/mT. The detection limit is defined as the field corresponding to a signal to noise ratio equal to 1. It is the minimum magnetic field that can be detected by the sensor. Using equation (5) the factor to normalize (multiply) the detection limit curves is $\sqrt{A.N}$, and it is validated by figure 3(d).

Thus normalization and correction factors (for a single pillar of area 1 μm²) for the noise, the sensitivity and the detectivity are important and are summarized in table 3. The measured performances (raw data) need to be corrected by the factor shown in the table 1 depending on the electronic circuit used, but also normalized by the number of pillars and by their area. Besides the detectivity normalized and corrected could also be extracted from the ratio of the corrected and normalized noise and sensitivity.

## IV. CONCLUSION

In this work, we presented a method to implement the correction due to the electronic circuit and to normalize the output noise and detection limit using parameters such as the number of MTJ pillars and pillar surface. We validated this method on TMR sensors by performing multiple measurements with various parameters and circuits configurations. This correction and normalization procedure can now be used to reliably compare the performances of TMR sensors from literature.

**Elmer Monteblanco** received the Ph.D. degree from the University Grenoble-Alpes, France, in 2014. He is currently a postdoc at SPEC, CEA Saclay. His research interests are spin electronics-based magnetic sensors, spintronics, nanomagnetism, and spin-dependent electronic transport in nanostructures. Founder of www.cientificos.pe.

**Aurélie Solignac** received the Ph.D. degree from the University of Pierre et Marie Curie, France, in 2012. She is currently a Research Engineer with SPEC, CEA Saclay. Her research interests are spin electronics-based magnetic sensors, nanomagnetism, and magnetic imaging.

**Julien Moulin** received the Ph.D. degree from the University of Paris Saclay, France, in 2020, on magnetic sensors development for magnetic microscopy.

**Chloé Chopin** received the Ph.D. degree from the University of Paris Saclay, France, in 2020, on magnetic sensors development for magnetophysiology applications

**Pierre Beliot** received the EE engineering degree from École supérieure d'électronique de l'Ouest, Angers-France, in 2000. He is currently working at Crivasense Technology and develops magnetoresistive sensors for automotive applications.

**Noémie Belin** received the engineering degree from INSA Rennes in 2008. She is currently working on the development of magnetic sensors at Crivasense Technologies.

**Paolo Campiglio** is currently working at Crivasense Technology and at Allegro Microsystems and develops magnetoresistive sensors for automotive applications

**Claude Fermon** received the Ph.D. degree from the University of ParisVI in 1986. He is a Senior Scientist with SPEC, CEA Saclay. His research interests are nanomagnetism, spin dynamics in nanometric objects, spin-dependent electronic transport in nanostructures, biomagnetism, and spin electronics based magnetic sensors.

**Myriam Pannetier-Lecoeur** received the Ph.D. degree from the University of Caen, France, in 1999. She is currently a Senior Scientist with SPEC, CEA Saclay. Her research interests are nanomagnetism, biomagnetism, and spin electronics-based magnetic sensors.